\documentclass{article}
\usepackage[T1]{fontenc} 
\usepackage[utf8]{inputenc} 
\usepackage{ismir,amsmath,cite,url}
\usepackage{graphicx}
\usepackage{color}
\usepackage{colortbl}
\usepackage{pifont}
\usepackage{multirow}
\usepackage{tabularx}
\newcolumntype{Y}{>{\centering\arraybackslash}X}
\usepackage{booktabs}
\usepackage{siunitx}
\usepackage{microtype}
\usepackage[capitalize]{cleveref}
\usepackage{savesym} 
\usepackage{musicography}
\savesymbol{meterC}
\usepackage[minimal]{leadsheets}
\useleadsheetslibraries{musicsymbols}
\usepackage{stmaryrd}

\usepackage{todonotes}

\usepackage{makecell}
\usepackage{xspace}
\def\ie{\textit{i.e.,}\xspace}
\def\eg{\textit{e.g.,}\xspace}
\def\wrt{w.r.t.\xspace}

\def\<#1>{\langle #1 \rangle}

\newcommand{\notes}{\ensuremath{\mathcal{\aleph}}}
\newcommand{\note}{\ensuremath{\mathcal{\eta}}}
\newcommand{\tpc}{\ensuremath{\mathcal{\tau}}}
\newcommand{\pc}{\ensuremath{p}}
\newcommand{\dur}{\ensuremath{d}}
\newcommand{\ks}{\ensuremath{\mathcal{\kappa}}}

\def\Music21{\textsf{Music21}\xspace}

\def\ps13{\textsf{\small ps13}\xspace}
\def\CIV{\textsf{\small CIV}\xspace}
\def\ourapp{\textsf{\small PKSpell}\xspace}

\def\Reduction#1#2#3#4{%
\mathrel{\raise1.0ex\hbox{%
\vtop{\ialign{##\crcr%
\raise0.0ex\hbox{$\hfil\scriptstyle{\ #1\ }\hfil$}\crcr%
\noalign{\nointerlineskip}%
\rightarrowfill\crcr%
\noalign{\nointerlineskip}%
\raise0.0ex\hbox{$\hfil\scriptstyle{\ #2\ }\hfil$}\crcr}}}{}^{#3}_{#4}}}
\def\Leduction#1#2#3#4{%
\mathrel{\raise1.0ex\hbox{%
\vtop{\ialign{##\crcr%
\raise0.0ex\hbox{$\hfil\scriptstyle{\ #1\ }\hfil$}\crcr%
\noalign{\nointerlineskip}%
\leftarrowfill\crcr%
\noalign{\nointerlineskip}%
$\hfil\scriptstyle{\ #2\ }\hfil$\crcr}}}{}^{#3}_{#4}}}
\def\hookReduction#1#2#3#4{%
\mathrel{\raise1.2ex\hbox{%
\vtop{\ialign{##\crcr%
\raise0.0ex\hbox{$\hfil\scriptstyle{\ #1\ }\hfil$}\crcr%
\noalign{\nointerlineskip}%
$\lhook\joinrel$\hspace{-0.35em}
\rightarrowfill\crcr%
\noalign{\nointerlineskip}%
$\hfil\scriptstyle{\ #2\ }\hfil$\crcr}}}{}^{#3}_{#4}}}
\def\hoookReduction#1#2#3#4{%
\lhook\joinrel\hspace{-0.50em}
\raise0.85ex\hbox{%
\vtop{\ialign{##\crcr%
\raise0.4ex\hbox{$\hfil\scriptstyle{\ #1\ }\hfil$}\crcr%
\noalign{\nointerlineskip}%
\rightarrowfill\crcr%
\noalign{\nointerlineskip}%
$\hfil\scriptstyle{\ #2\ }\hfil$\crcr}}}{}^{#3}_{#4}}

\def\frew#1#2#3#4#5#6#7#8{
\setbox0=\hbox{$#6 #7 #1 #8$}%
\setbox1=\hbox{$#6 #7 #2 #8$}%
\ifdim \wd0>\wd1 \rlap{\rlap{\hbox to \wd0{#5}}%
                            {\hbox to\wd0{\hfil\lower #3\box1\relax\hfil}}}{\raise #4\box0}%
\else \rlap{\rlap{\hbox to \wd1{#5}}{\hbox to\wd1{\hfil\raise #4\box0\relax\hfil}}}{\lower #3\box1}%
\fi
}

\usepackage{listings}
\usepackage{lineno}

\title{PKSPELL: Data-Driven Pitch Spelling and \\ Key Signature Estimation}

\threeauthors
  {Francesco Foscarin} {CNAM Paris \\ {\tt francesco.foscarin@cnam.fr}}
  {Nicolas Audebert } {CNAM Paris \\ {\tt nicolas.audebert@cnam.fr}}
  {Raphaël Fournier S'niehotta } {CNAM Paris \\ {\tt fournier@cnam.fr}}
  
\def\authorname{F. Foscarin, N. Audebert, and R. Fournier S'niehotta}

\begin{document}

\maketitle
%
\begin{abstract}
We present \ourapp: a data-driven approach for the joint estimation of pitch spelling and key signatures from MIDI files. Both elements are fundamental for the production of a full-fledged musical score and facilitate many MIR tasks such as harmonic analysis, section identification, melodic similarity, and search in a digital music library.

We design a deep recurrent neural network model that only requires information readily available in all kinds of MIDI files, including performances, or other symbolic encodings. We release a model trained on the ASAP dataset. Our system can be used with these pre-trained parameters and is easy to integrate into a MIR pipeline. We also propose a data augmentation procedure that helps re-training on small datasets.

\ourapp achieves strong key signature estimation performance on a challenging dataset. Most importantly, this model establishes a new state-of-the-art performance on the MuseData pitch spelling dataset without retraining.
\end{abstract}
%
\section{Introduction}
\label{sec:introduction}
Music listening is a complex cognitive process that involves the organization of sound events in time and frequency structures.
This process creates patterns of thought that depend either on general principles of cognitive psychology or on prior knowledge and common practices of the relevant musical style system~\cite{krumhansl1990cognitive}.
As far as tonal music is concerned, pitches are related and arranged to create a complex system of perceived stability and instability that gravitates around a tonal center, usually identified by a scale or a chord~\cite{diatonic1997}.
In this paper, we focus on two aspects of the tonal framework: pitch spelling and key signature.

\subsection{Pitch Spelling}
\begin{figure}
 \centerline{
 \includegraphics[width=0.96\columnwidth]{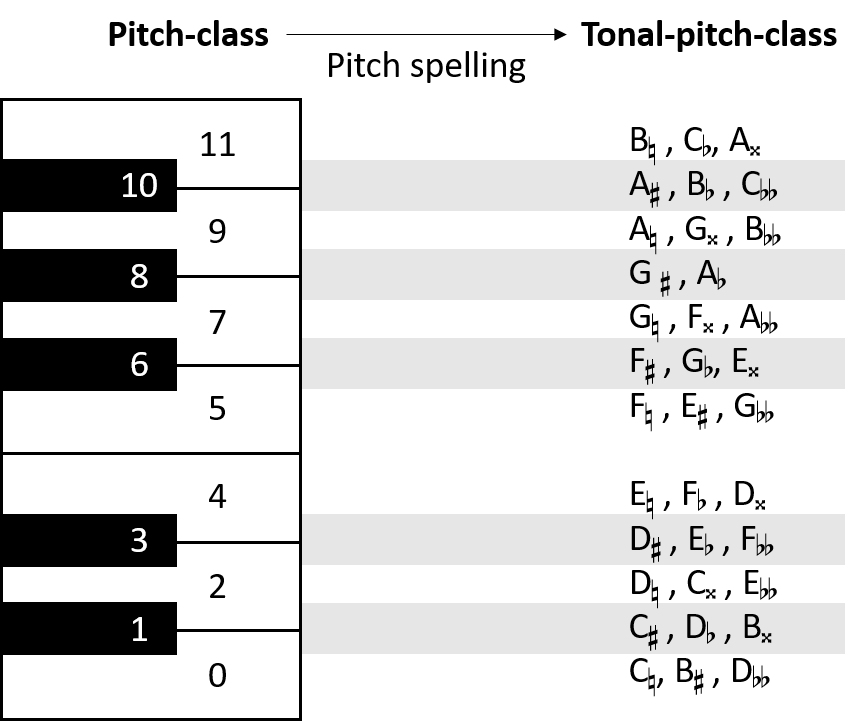}}
 \caption{Pitch-classes and corresponding tonal-pitch-classes.}
 \label{fig:pitch_class}
\end{figure}

In tonal music, the classification of pitches in 12 \textit{pitch-classes}, each one uniquely identifying a set of frequencies that are $n$ octaves apart, is enriched with some tonal information and creates the higher-level representation of \textit{tonal-pitch-classes}. Each \textit{tonal-pitch-class} consists of a \textit{diatonic name} in $[C,D,E,F,G,A,B]$ and an \textit{accidental} among double-flat (\musDoubleFlat), flat (\musFlat), natural(\musNatural), sharp (\musSharp), double-sharp (\musDoubleSharp).\footnote{More accidental types, like triple sharps could exist in theory, but they are not used in common music notation.}
For every pitch-class, there are multiple corresponding tonal-pitch-classes, called \textit{enharmonic equivalents} (see \cref{fig:pitch_class}). The task of choosing a single name between the possible enharmonic equivalents is called \textit{pitch spelling}.

\subsection{Key Signature}\label{sec:key_signature}
\begin{figure}
 \centerline{
 \includegraphics[trim=0 17 0 10, clip, width=0.96\columnwidth]{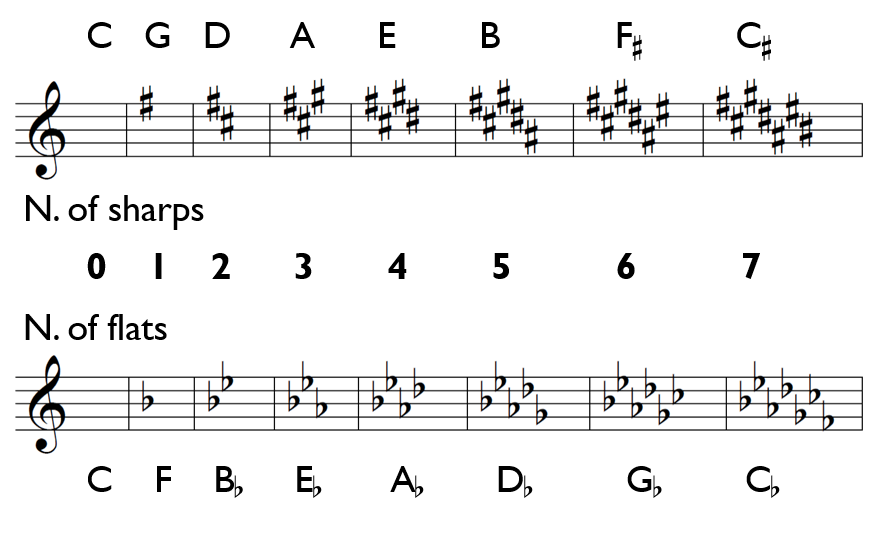}}
 \caption{Corresponding key signatures representation in a musical score, tonal-pitch-class and number of accidentals.}
 \label{fig:key_signatures}
\end{figure}

\textit{Key signatures} inform about the tonal center of the piece and are commonly identified in music by a certain number of sharps or flats (whose positions are fixed) or by a tonal-pitch-class (see \cref{fig:key_signatures}).
Combined with pitch spelling, they also improve readability, according to the principle of \textit{notation parsimony}~\cite{cambouropoulos2003pitch}, \ie the minimization of the number of symbols (accidentals) displayed in the score.

Tightly related to the key signature, the \textit{key} also adds information about the mode (major or minor). Keys are considered in literature either on a global scale (\ie one \textit{global key} for one piece) or a local scale (\ie multiple \textit{local keys} for one piece) resulting from modulations and tonicizations (see~\cite{lopez2020local} for a complete description of these concepts).
A formalization of the relation between local keys, global key, and key signatures is missing in the literature and it is not the goal of this paper to have a musicological discussion on this topic.
As a first approximation, the key signature can be considered a global feature of the piece. It occasionally changes if the tonal center shifts significantly from its previous position, although it does not move as much as the local key.
A possible motivation is that, since the key signature is employed in music notation, the principle of parsimony applies and it ``smoothes'' short key changes.

\subsection{Pitch Spelling and Key Signature Estimation}
Multiple reasons motivate systems that perform pitch spelling and key signature estimation. Such information is necessary for the creation of full-fledged musical scores, \eg as the last step of automatic music transcription or music generation systems. Furthermore, those elements are employed for other MIR tasks: key signature annotations are useful for music search and section identification, and pitch spelling facilitates harmonic analysis and melodic pattern matching~\cite{cambouropoulos2003pitch,temperley2001cognition}. 
However, they are not represented in ``low-level'' formats, such as audio and MIDI files\footnote{Albeit key signatures can be encoded in a MIDI file as metadata, they are often absent.}, which constitute most of the available digital musical encoding.

\subsection{Related Works}
Several authors have developed automatic systems that perform pitch spelling from MIDI data. Most of them consist in algorithmic approaches based on musicological insights only, without any learning. Some authors think the selection of the appropriate tonal-pitch-class is a function of the local key~\cite{chew2005pitch_spelling, meredith2006ps13} or a combination of the local key and voice-leading information~\cite{longuet_higgins1971_bach,temperley2001cognition}, \ie the temporal evolution of notes within each voice.
Others base pitch spelling on a type of interval optimizations~\cite{wetherfield_2020_pitchspelling}, on the principle of notational parsimony, or a combination of the two criteria\cite{cambouropoulos2003pitch}.
Another formulation of the problem as a generative probabilistic model is proposed by Teodoru et Raphael~\cite{teodoru2007pitch}. They model voices with independent Markov chains conditioned on a hidden state that contains the local key information.

For evaluation purposes, the early algorithmic approaches were compared by Meredith~\cite{meredith2006ps13} on a dataset of 216 pieces by 8 baroque and classical composers. The highest accuracy at the time was obtained by Meredith's \ps13 algorithm, especially when small temporal variations were introduced to simulate a MIDI file generated from a performance. 
Teodoru and Raphael~\cite{teodoru2007pitch} also tested their approach on the same dataset (without tempo deviation) with high accuracy. More recent works~\cite{honingh2009lattice,wetherfield_2020_pitchspelling} did not increase the total accuracy.
In this paper, we compare our results with Meredith's \ps13 algorithm and Teodoru and Raphael's approach with conditional independent voices (\CIV).

While those approaches yield a very high accuracy ($\geq 99$\% of notes are correctly spelled), they are not designed to be easily employed in larger MIR pipelines.
\ps13 has 3 different outputs, whose tonal pitch classes are transposed by diminished seconds (\eg one piece in C\musSharp{}, one in D\musFlat{} and one in B\musDoubleSharp{}) and does not indicate how to select the ``best'' version.
Both \ps13 and \CIV parameters are set by hand\footnote{\CIV could be trained, but the authors report that the results are worse than with handset parameters.}, thus making them difficult to generalize across composers and datasets. Moreover, the code of \CIV is not publicly available.
Finally, both approaches do not provide key signatures, so they cannot be used alone to produce a complete pitch encoding for a musical score.
Other pitch spelling systems are implemented in music notation software (\eg Finale, MuseScore) but no information about their functioning mechanism and performance is available and they require key signatures as input.

Little attention has been given in the literature to the {\em key signature estimation} problem from MIDI files, in favor of the related task of key estimation~\cite{chew2002spiral,temperley2008key,albrecht2013largecorpora,lopez2019key} (see \cref{sec:key_signature}). A direct comparison of our results with other approaches is therefore not possible, but we put our results into perspective by comparing them with the state-of-the-art approach for global key estimation of López et al.~\cite{lopez2019key}.

\subsection{Our approach}
We propose the \ourapp system for jointly estimating pitch spelling and key signature. It yields high accuracy, is easy to integrate into a  MIR pipeline, and works on any kind of MIDI file (including MIDI generated from human performance) or other symbolic encodings. Trained on the ASAP dataset~\cite{foscarin2020asap}, \ourapp achieves strong key signature estimation performance on the Albrecht dataset~\cite{albrecht2013largecorpora} and establishes a new state-of-the-art pitch spelling performance on the MuseData dataset~\cite{meredith2006ps13} without retraining. Implementation and pre-trained model are publicly available\footnote{See \url{https://github.com/fosfrancesco/pkspell}}.

We design a deep learning approach based on recurrent neural networks (RNN) that can model correlations in input sequences of variable lengths~\cite{SHERSTINSKY2020rnn}. We use a dedicated network structure inspired by musicological considerations on the relation between local keys, pitch spelling, and key signatures (details in \cref{sec:rnnmodel}). Our system models each input note with a pair (pitch-class, duration) that does not require high-level temporal information such as downbeat and time signature or voice separation to be produced. With the proposed dedicated preprocessing of the note durations and data augmentation procedure (see \cref{sec:input,sec:augmentation}), our approach can generalize to different tempos, time signatures, and key signatures. This makes it possible to train our model on a small-size dataset of musical scores. Moreover, by cross-validating on a separated dataset, we show that with the pre-trained parameters, our system can correctly handle a variety of different tonal pieces. Finally, our multi-task approach to pitch spelling and key signature estimation increases the performance on both tasks.

\section{Method}
\label{sec:method}

The pitch spelling and key signature information are the results of the relations between different notes. We employ a recurrent neural network architecture, which can learn all those relations for pieces of different lengths. This class of models employs a big number of parameters to correctly learn the input-output function. Our choice of input/output representation helps to keep their number as low as possible, and the data augmentation procedure we propose allows us to learn them from a relatively small dataset.

\subsection{Input and output formats}\label{sec:input}
Multiple approaches have been proposed for the problem of transforming a MIDI file into a sequential representation~\cite{micchi2020not,oore2020time,thickstun2018coupled}.
Since we target all kinds of MIDI inputs (see~\cite{midi_spec} for a description of different MIDI formats), we cannot assume to have information such as voice separation, time signature, downbeat, and beat positions. We employ a simple note-based representation that was proposed by Lopez~\cite{lopez2019key}: a piece is modeled as a sequence of notes $\notes = \{\note_1,\note_2,...,\note_N\}$, ordered according to the temporal position of their onsets. If multiple notes have the same onset position, we take the notes in low-to-high order.\footnote{ However, the ordering choice does not impact the results.}
For each note $\note$, we consider two features: pitch-class $\pc[\note]$ and duration $\dur[\note]$. The usage of note durations is common in key estimation approaches (\eg \cite{krumhansl1990cognitive, chew2002spiral}), and stems from the musicological intuition that longer notes have a stronger impact than shorter notes in defining the tonal context.

\begin{figure}
 \centerline{
 \includegraphics[width=0.999\columnwidth]{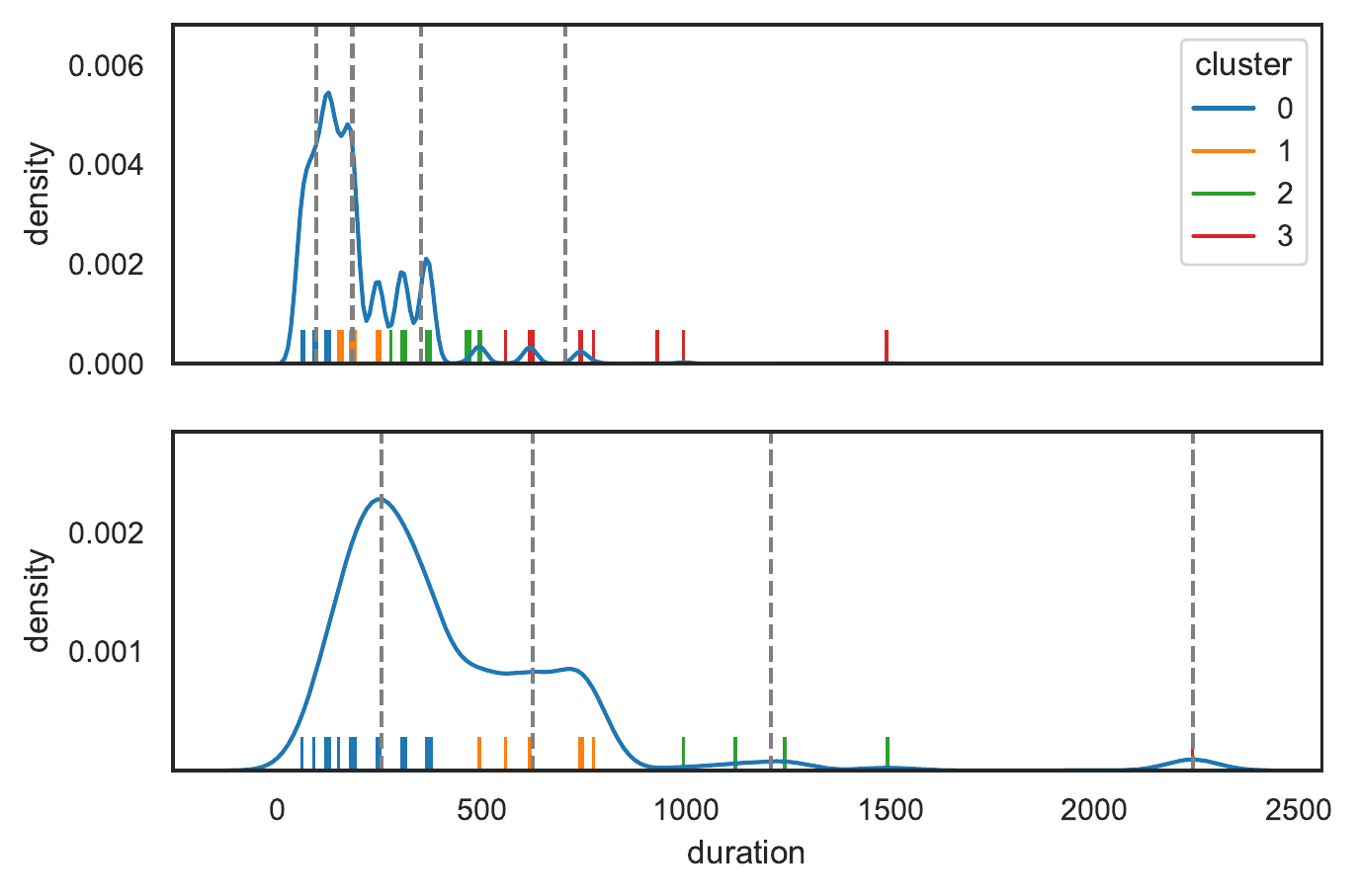}}
 \caption{One-dimensional clustering of note durations for 2 pieces expressed in milliseconds. Each vertical line represent a duration and dotted grey lines are cluster centroids. The piece at top is faster than the piece at the bottom. To facilitate the understanding, the kernel density estimation of the note durations is also displayed. }
 \label{fig:kmeans}
\end{figure}

To group all possible note durations in a limited number of classes, we run an (unsupervised) 1-dimensional {\em k-means} algorithm for all note durations (see \cref{fig:kmeans}). We use the dynamic programming algorithm presented by Gronlund et al.~\cite{gronlund2017fast} that has complexity $O(kN + N log N)$, for $k$ classes and a sequence of length $N$. We select $k=4$.
This classification cannot be considered more than a rough indication of relative duration and it is not meant to precisely identify beats, downbeats, and other metrical information. Moreover, it will show its limits if there are tempo changes or time signature changes inside the piece. Yet, we found that it improved our model results at a small computational cost and, compared to other possible approaches, \eg quantization to some discrete durations, it has the advantage of generalizing to different tempos and time signatures.

The output is a sequence of tonal-pitch-classes $\tpc[\note]$ (among the 35 in \cref{fig:pitch_class}) and a sequence of key signatures $\ks[\note]$ (among the 15 in \cref{fig:key_signatures}), one for each note $\note$ in input. Computing a key signature for each note might seem excessive, since we may expect the key signature to change only a few times during a piece. However, this is necessary as we do not know in advance how many changes there are in a piece, nor do we have precise metrical information. 

\subsection{Data augmentation}
\label{sec:augmentation}
Studies in cognitive psychology~\cite{dowling1978scale} have proved that listeners perceive as identical two pieces if all their notes are transposed by the same \textit{interval}.
It is common to use this property of music perception to augment a dataset by transposition~\cite{micchi2020not,chen2019harmony}.
For our goals, we must transpose pitch-classes (input) and tonal-pitch-classes and key signatures (output) to have a correct ground truth for training. 

\begin{figure}
 \centerline{
 \includegraphics[width=0.96\columnwidth]{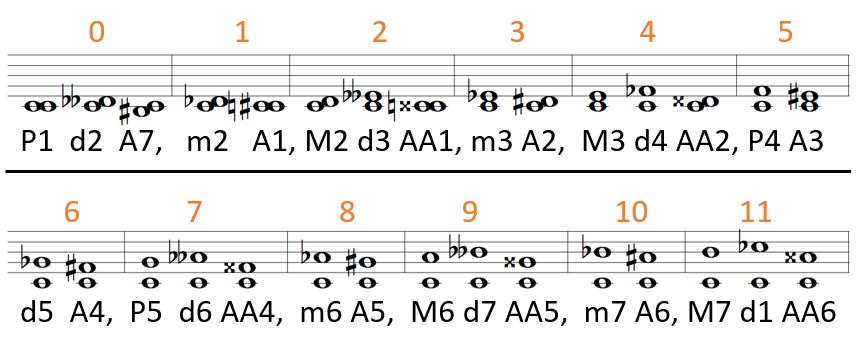}}
 \caption{Chromatic intervals (orange), corresponding diatonic intervals (black) and examples from the tonal-pitch-class $C$. The abbreviation for diatonic types are ``P'': perfect, ``M'': major, ``m'':minor, ``A'': augmented, ``d'': diminished, ``AA'': doubly augmented.}
 \label{fig:intervals}
\end{figure}

The possible transpositions of tonal-pitch-classes and key signatures move through a \textit{spiral of fifth}s~\cite{chew2002spiral}. When reported to the same octave, each transposition can be identified with a \textit{diatonic interval}, notated with an interval number and type, \eg augmented 4th, perfect 5th, minor 2nd. For pitch-classes, instead, we are limited to 12 \textit{chromatic intervals} that can be simply identified by integers.
In \cref{fig:intervals} we report the most common diatonic intervals (the ones closer to the center of the spiral of fifths) associated with their respective chromatic interval.

Since our goal is to learn an input-output mapping, we cannot accept multiple sequences of tonal-pitch-classes that correspond to the same sequence of pitch-class. That means that we need to select only one diatonic interval for every chromatic interval.
To do so, we use a heuristic based on the principle of parsimony, \ie we choose the diatonic transposition which generates the set of tonal pitch classes with the lowest number of accidentals.
For each piece and each chromatic transposition, we transpose the tonal pitch class by all corresponding diatonic intervals, then we discard transpositions that contain non-accepted accidentals (\eg triple sharps). Finally, we select as the ``valid'' diatonic transposition the one with the smallest count of accidentals, where \musDoubleFlat{} and \musDoubleSharp{} count as 2, and \musSharp{} and \musFlat{} count as~1. This allows us to produce up to 11 variants for each piece, although the number is lower if all diatonic intervals for a certain chromatic interval generate a discarded transposition.

\subsection{RNN Architecture}\label{sec:rnnmodel}
With our input and output choice, the pitch spelling and key signature estimation problems for one piece can be seen as a sequence of multi-task classification problems. Our goal is to assign to each input (\ie pitch-class $\pc$ + duration $\dur$) two labels: one among the available tonal-pitch-classes $\tpc$ and one among the key signatures $\ks$.
Formally, for a piece $C$ and each note $\note \in \notes_C$ we seek to learn the function $F: x[\note] \rightarrow y[\note]$, where $x[\note] = \pc[\note], \dur[\note]$ is our input and $y[\note]= \tpc[\note],\ks[\note]$ is our output. 

We want the output for each note to be dependent on the notes around it, and the length of our sequences (\ie the number of notes in a piece) is not fixed. We select as the core of our model a recurrent neural network (RNN), that can store information about the ``context'' in its internal state for variable-length sequences of inputs.
Such a model can be trained in a supervised fashion on an annotated dataset that contains for each piece a sequence of pairs $\{(x[\note], y[\note]), \forall \note \in \notes_C\}$. We optimize the parameters $\theta$ \wrt a classification loss function $\mathcal{L}$:
$$\theta^* = \arg\min_\theta \sum_{C} \sum_{\note \in \notes_C} \mathcal{L}(F_\theta(x[\note]), y[\note])$$
We select $\mathcal{L}$ as the sum of two cross-entropy-loss functions~\cite{murphy2012machine}, one for the key signature and one for the tonal-pitch-class. Since $\mathcal{L}$ is differentiable, the model parameters can be optimized using Stochastic Gradient Descent (SGD).

\begin{figure}
 \centerline{
 \includegraphics[width=0.95\columnwidth]{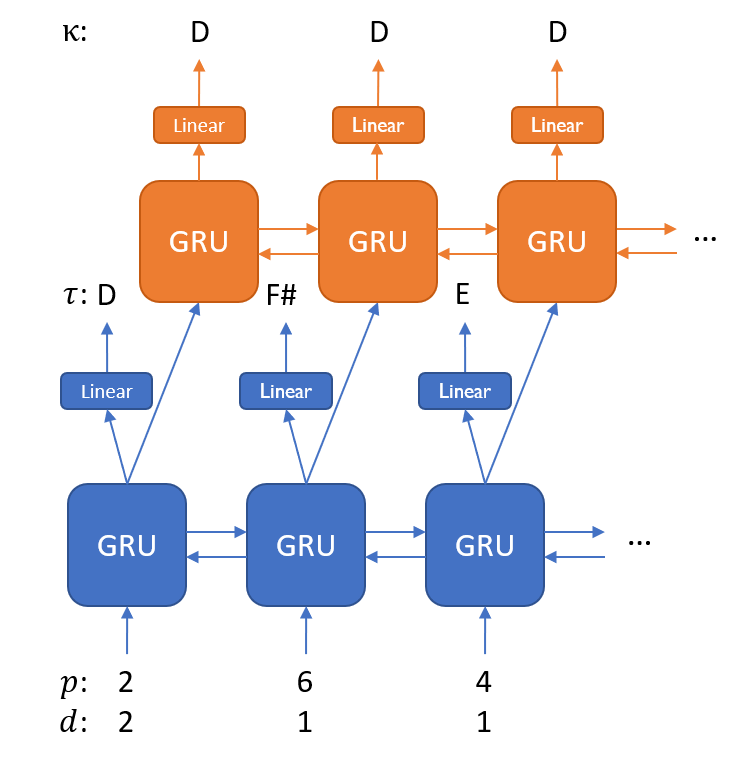}}
 \caption{Model architecture. For each pitch-class $\pc$ and duration $\dur$ in input, a tonal-pitch-class $\tpc$ and a key signature $\ks$ are produced.}
 \label{fig:architecture}
\end{figure}

We use a bidirectional RNN to tie the output at a certain note to the past and future inputs. From a musicological standpoint, ``seeing the future'' is useful \eg to know where a certain note will \textit{resolve}.
As shown in \cref{fig:architecture}, the model has two recurrent layers, each one coupled with a linear layer. The first produces the tonal-pitch-classes and the second produces the key signatures. 
This architecture is based on the musicological hypothesis that pitch spelling depends on the local key~\cite{chew2005pitch_spelling,meredith2006ps13} and the key signature is a ``smoothed'' version of the local key. Assuming the first layer encodes information about the local key, we aggregate this information in the second recurrent layer to infer the key signature. 

According to the principle of \textit{multi-task learning}~\cite{collobert2008unified,deng2013new}, jointly producing two outputs allows the model to learn shared representations, thus enabling a better generalization on both our original tasks.

\section{Experiments and Results}
\label{sec:results}

We train our model on the pieces from the ASAP dataset~\cite{foscarin2020asap}. 
The dataset provides 222 compositions from several composers over two centuries: Bach, Beethoven, Chopin, Schubert, Haydn, Liszt, Schumann, Mozart, Rachmaninoff, Ravel, Debussy, Scriabin, Glinka, Brahms, Prokofiev, Balakirev. We remove two pieces because they overlap with the dataset that we use for pitch spelling evaluation (see \cref{sec:main_results}). 
All the information we need for training can be easily extracted from digitally encoded musical scores, \ie MusicXML scores for the ASAP dataset.
Such scores contain note durations, tonal-pitch-classes, and key signatures. From tonal-pitch-classes it is straightforward to produce pitch-classes using the function in \cref{fig:pitch_class}.
Duration and pitch-classes are then encoded as one-hot vectors, concatenated to a single vector, and fed into our model.
Training on real MIDI performances could also be performed as long as a note-wise score to performance alignment is available. Unfortunately, ASAP does not provide this alignment, and other datasets that provide such information are considerably smaller.

To evaluate the benchmarked approaches, we use the accuracy, \ie the percentage of correctly inferred symbols. We also use the error rate ($1-\textrm{accuracy}$)  to improve the readability of some results.

\subsection{Hyperparameter search}
Our model has five major hyperparameters to consider:  type of optimizer, learning rate, batch size, type of recurrent cell, and number of parameters (number of layers and hidden state dimension for each RNN). To find the best combination, we perform a grid search by training our model on 85\% of ASAP (187 pieces, 2033 after data augmentation) and validating on the remaining 15\% (33 pieces).

\textbf{Optimizer}. 
We compare various optimizers from the deep learning literature, including gradient descent algorithms with adaptive learning rates or momentum. We find no significant difference in accuracy; though Adam\cite{kingma_adam:_2015} is easier to tune and converges faster than traditional SGD.

\textbf{Learning rate}.
All networks are trained for 40 epochs, \ie 40 passes over the whole augmented training set.
We find that a starting learning rate of 0.01, divided by 10 after 20 epochs, allows for fast and robust convergence.

\textbf{Batch size}.
We feed mini-batches of sequences in parallel to our model. 
For batches larger than $32$, changing the batch size has no impact on performance. Larger batches tend to increase convergence speed but with a higher GPU memory consumption; thus we settle for a batch size of 32.

\textbf{Recurrent cell}.
We compare LSTM\cite{10.1162/neco.1997.9.8.1735} and GRU\cite{cho-etal-2014-learning} cells. While the accuracy is similar with the two cells, we find that GRU cells are faster and use less memory.

\textbf{Number of parameters}.
There are two ways to increase the number of parameters in the RNN: widen it by increasing the hidden dimension, or deepen it by stacking more layers. 
More parameters generally entail better performance on the train set but not necessarily on the test set due to overfitting. Dropout\cite{JMLR:v15:srivastava14a} is required to alleviate overfitting for a depth greater than 1 or a hidden dimension higher than $\approx 200$. 
For each of the two RNN in our final model, we set depth 1 and hidden dimension 300 (150 in each direction).

\subsection{Main results}\label{sec:main_results}
We evaluate the pitch spelling and key signature estimation results separately.

For pitch spelling, we compare with \ps13 and \CIV on the MuseData dataset proposed by Meredith~\cite{meredith2006ps13}. It consists of 216 pieces (\num{195972} notes) from 8 classical composers (see \cref{tab:evaluation}) and is also available in a ``noisy'' version, where some noise is artificially introduced in the onset and offset positions to roughly simulate human performances. There are no pieces in common with our training dataset.
It is worth noting that both \ps13 and \CIV tune their parameters on the test data; this can induce overfitting and thus an optimistic estimate of the system's performance. We may notice, for example, that the pieces used to train \CIV (all Beethoven's, one-third of Haydn's, and one-third of Mozart's), correspond to the pieces in which \CIV has the highest performance compared to our system.
Moreover, both \ps13 and our system are evaluated on the ``noisy'' dataset, while \CIV is evaluated on the quantized dataset.
We report our evaluation results in \cref{tab:evaluation}. \ourapp establishes new state-of-the-art performances on the pitch spelling task. It does 256 errors,   around 25\% less than \CIV (343) and 75\% less than \ps13 (1064).
The learned model has an excellent generalization, going close to zero error if the pieces have a unique strong tonal center (\eg Corelli, Handel, Telemann, Vivaldi). More difficult are instances when multiple modulations happen during a short time span (\eg Bach chorales), and especially around abrupt key signature changes (\eg Beethoven). The error count for Haydn is particularly high because of the piece Symphony No. 100 in G major, where, at measure 166 there is a sudden change from D\musFlat{} major to C\musSharp{}, to make the music easier to read with fewer accidentals.
This kind of enharmonic key change is rare in music and our system fails to detect it.\footnote{\ps13 manually transposes half of this piece before evaluation. There is no indication of its treatment by \CIV.}
It is worth noticing that, while our model allows a pitch-class to be classified in a non-correspondent tonal-pitch-class (\eg 2$\rightarrow$G\musSharp), this kind of error is completely absent in our results. Our model learns very easily to perform the mapping of \cref{fig:pitch_class}, especially when using the augmented dataset.

\begin{table*}
\begin{center}
\begin{tabularx}{\textwidth}{lYYYYYYYYr}
    \toprule
     & Bach  & Beethoven & Corelli & Handel & Haydn & Mozart & Telemann & Vivaldi &   \textbf{Total} \\
     \midrule
\ps13~\cite{meredith2006ps13} & \makecell{0.17\% \\ (42)}   & \makecell{1.41\% \\ (345)}     & \makecell{0.04\% \\ (11)}     & \makecell{0.26\% \\ (68)}    & \makecell{0.94\% \\ (220)}   & \makecell{0.23\% \\ (282)}   & \makecell{0.34\% \\ (82)}     & \makecell{0.39\% \\ (114)}     & \makecell{0.59\% \\ (1064)}  \\
\CIV~\cite{teodoru2007pitch}  & \makecell{0.10\% \\ (25)} &  \makecell{\textbf{0.10\%} \\ \textbf{(25)*}} & \makecell{0.08\% \\ (20)} & \makecell{0.02\% \\ (6)} & \makecell{\textbf{0.45\%} \\ \textbf{(110)*}} & \makecell{0.33\% \\ (79)*} & \makecell{0.05\% \\ (13)} & \makecell{0.27\% \\ (65)} & \makecell{0.18\% \\ (343)}\\

    \midrule

\ourapp  & \makecell{\textbf{0.08\%} \\ \textbf{(19)}} & \makecell{0.23\% \\ (56)} & \makecell{\textbf{0.02\%} \\\textbf{(5)}} & \makecell{\textbf{0.02\%} \\ \textbf{(5)}} & \makecell{0.49\% \\ (121)} & \makecell{\textbf{0.14\%} \\ \textbf{(35)}} & \makecell{\textbf{0.02\%} \\ \textbf{(4)}} & \makecell{\textbf{0.04\%} \\ \textbf{(11)}} & \makecell{\textbf{0.13\%} \\ \textbf{(256)}}\\
    \bottomrule
\end{tabularx}
\end{center}
\caption{Error rate 
and the number of errors (between parentheses) for different composers in Meredith's Musedata pitch spelling dataset~\cite{meredith2006ps13}. For \ps13 and \CIV, we took these values from the respective papers. The symbol * marks the results for the composers used to set the parameters for \CIV. The best result for each composer is highlighted in bold.}
\label{tab:evaluation}
\end{table*}

While a direct evaluation of the key signature estimation is not possible due to the lack of other approaches targeting this task, we put into perspective our results by evaluating our system on the Albrecht dataset~\cite{albrecht2013largecorpora} and comparing it with the state-of-the-art results for the global key signature estimation of López et al.~\cite{lopez2019key}. 
Our task can be considered easier because we do not need to separate major keys from minor keys; however, while we are considering all enharmonic key signatures (\eg D\musFlat{} with five flats is considered equivalent to C\musSharp{} with seven sharps), López is flattening all enharmonic versions of a key to the same class.
After removing the pieces in common with ASAP, we are left with 932 pieces. We correctly classify 97\% of the global key signatures. For comparison, López et al. correctly classify 94\% of the keys. Of the 29 misspelled pieces, eight are mapped to enharmonically equivalent key signatures, in particular, there is confusion between the C\musSharp{} and D\musFlat{} and between F\musSharp{} and G\musFlat{}. The remaining 21 wrong estimations are off by one accidental, \ie the predicted key signature is the relative 4th or 5th of the true key signature.

\subsection{Ablation Studies}
\begin{figure}
 \centerline{
 \includegraphics[width=0.999\columnwidth]{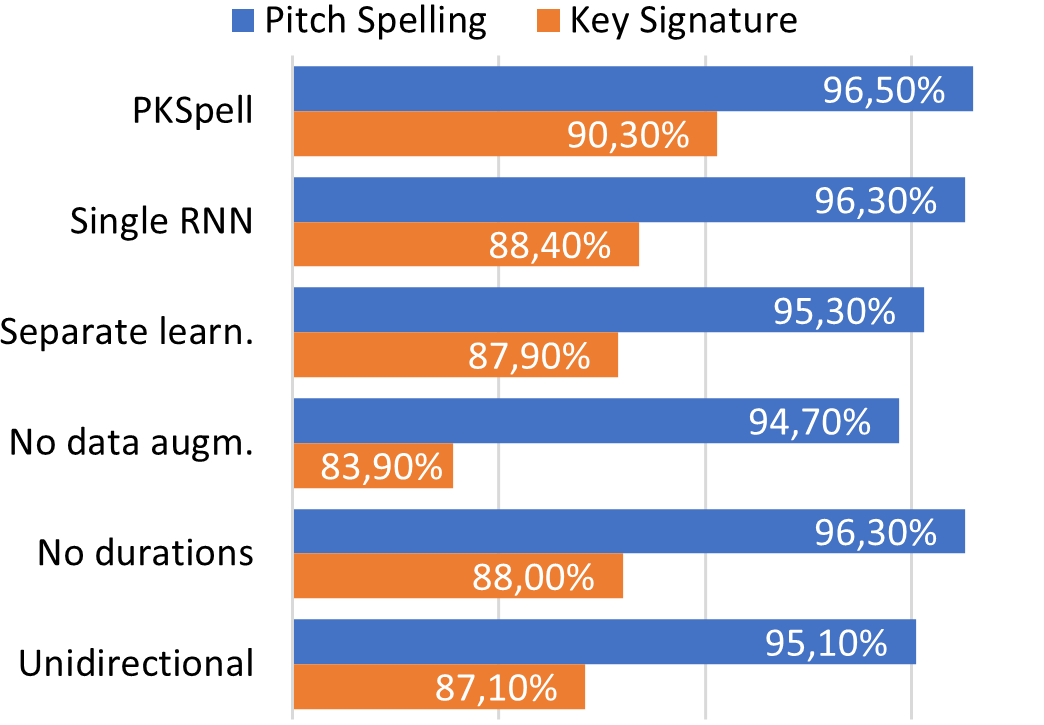}}
 \caption{Accuracies for the ablation tests. Results are reported on the validation set of ASAP (33 pieces).}
 \label{fig:ablation}
\end{figure}

We perform several ablation studies to understand how our design choices impact the model performance. For each experiment, we remove one element from \ourapp and see how this affects the model performance. If the element is useful, we expect a reduction in the accuracy (see \cref{fig:ablation}).

\textbf{Single RNN}. As previously introduced, \ourapp uses two recurrent layers, one for pitch spelling and another for key signature estimation. 
To evaluate this idea, we build a model that has a single recurrent layer for both tasks.
A detailed analysis shows that one-layer-\ourapp slightly outperforms the two-layer model when the tonal center is very stable. However, the two-layer model majorly improves the results for more pieces with modulations and key changes. On the ensemble of considered composers, the usage of two separate RNN layers boosts the accuracy, especially for key signature estimation (+3\%).

\textbf{Separate learning}. Multi-task learning can help leverage domain-specific information contained in related but different tasks. This is the case in our system, as the accuracy for both PS and KS estimation improves when trained jointly compared to two distinct RNNs used separately.

\textbf{No data augmentation}. In theory, more data should improve the generalization of the model, provided that the augmented samples are representative of future observations. In our case, data augmentation provides a significant accuracy boost, especially for key signature estimation. 

\textbf{No durations}. We use the duration of notes as an input feature for our model. While this is common in key estimation systems, multiple approaches to pitch spelling in the literature do not use this information \cite{meredith2006ps13,cambouropoulos2003pitch}. We find that durations improve our results, especially for KS estimation.

\textbf{Unidirectional RNN}. The RNN layers we consider can process a sequence either in one direction (usually left-to-right, LTR) or in a bidirectional manner, both LTR and RTL.
We expect the bidirectional processing to perform better since the model also ``sees'' future notes. In our ablation experiment, both PS and KS accuracies increase by more than 1 point by using both directions (we keep the same number of parameters by dividing the hidden dimension by 2 when we use the bidirectional model). Note, however, that the unidirectional LTR model still works reasonably well and could be used for a real-time version of our system.

\section{Conclusions and perspectives}
\label{sec:disconclu}

We introduce \ourapp, a novel system for joint pitch spelling and key signature estimation that reaches new state-of-the-art performances on pitch spelling, is easy to integrate into a MIR pipeline, and works on any MIDI file, including human performances. 
To reach this goal, we perform multi-task learning with an RNN model inspired by musicological insights. We propose a data augmentation procedure and a preprocessing of note durations to generalize to different transpositions, tempos, and time signatures.  We consider a minimal set of inputs (pitch-classes and note durations) that do not require high-level information such as time signature, voice separation, and downbeat position.
The pre-trained model we provide can be used ``as-is'' and offers good performance for classical music of different centuries.
Thanks to the ablation study, we directly observe the impact of the design choices of our system.

Possible future work concerns the evaluation of PKSpell on different styles of tonal music \eg pop, folk, and jazz. We expect it to perform well on pop and folk due to their low harmonic complexity. Jazz can be more challenging because of the extensive use of chord extensions and alterations. It is also interesting  to study how non-strictly-tonal music in the ASAP dataset (\eg Debussy) impacts the training of our model.
While concepts such as key signature and pitch spelling are less significant for non-tonal music, they are still used to write musical scores. Since 
\ourapp is not based 
on strong tonal principles, we expect it to be able to learn those rules as long as some coherent rules exist for pitch spelling and a large enough dataset is provided. 
A more in-depth analysis can be performed on the treatment of rhythmical information, by considering different ways to group duration values and by varying the number of groups.
Other improvements may be done on the model architecture: the state of the art for sequence-to-sequence problems has shifted toward recurrent attentional models and transformers. However, the length of the sequences we are considering (more than \num{15000} notes for certain pieces) poses a considerable challenge for full attentional mechanisms, whose memory requirements are quadratic with the input sequence length.
Models that scale linearly have been recently proposed~\cite{zaheer2020bigbird,xiong2021nystromformer} and could be a solution for this problem.
Finally, it would be interesting to employ our model to perform local and global key estimation. 

\section{Acknowledgments}
We thank David Meredith for releasing the MuseData corpus and ps13 Lisp implementation. Additional thanks go to Tim Bradshaw for help running ps13 on current systems.

\bibliography{ISMIRtemplate}

\begin{thebibliography}{10}
\providecommand{\url}[1]{#1}
\csname url@samestyle\endcsname
\providecommand{\newblock}{\relax}
\providecommand{\bibinfo}[2]{#2}
\providecommand{\BIBentrySTDinterwordspacing}{\spaceskip=0pt\relax}
\providecommand{\BIBentryALTinterwordstretchfactor}{4}
\providecommand{\BIBentryALTinterwordspacing}{\spaceskip=\fontdimen2\font plus
\BIBentryALTinterwordstretchfactor\fontdimen3\font minus
  \fontdimen4\font\relax}
\providecommand{\BIBforeignlanguage}[2]{{%
\expandafter\ifx\csname l@#1\endcsname\relax
\typeout{** WARNING: IEEEtran.bst: No hyphenation pattern has been}%
\typeout{** loaded for the language `#1'. Using the pattern for}%
\typeout{** the default language instead.}%
\else
\language=\csname l@#1\endcsname
\fi
#2}}
\providecommand{\BIBdecl}{\relax}
\BIBdecl

\bibitem{krumhansl1990cognitive}
C.~L. Krumhansl, \emph{Cognitive foundations of musical pitch}.\hskip 1em plus
  0.5em minus 0.4em\relax Oxford University Press, 1990.

\bibitem{diatonic1997}
R.~Van~Egmond and D.~Butler, ``Diatonic connotations of pitch-class sets,''
  \emph{Music Perception}, vol.~15, no.~1, pp. 1--29, 1997.

\bibitem{cambouropoulos2003pitch}
E.~Cambouropoulos, ``Pitch spelling: A computational model,'' \emph{Music
  Perception}, vol.~20, no.~4, pp. 411--429, 2003.

\bibitem{lopez2020local}
N.~N. L{\'o}pez, L.~Feisthauer, F.~Lev{\'e}, and I.~Fujinaga, ``On local keys,
  modulations, and tonicizations,'' in \emph{Digital Libraries for Musicology
  (DLfM)}, 2020.

\bibitem{temperley2001cognition}
D.~Temperley, \emph{The cognition of basic musical structures}.\hskip 1em plus
  0.5em minus 0.4em\relax MIT {Press}, 2001.

\bibitem{chew2005pitch_spelling}
E.~Chew and Y.-C. Chen, ``Real-time pitch spelling using the spiral array,''
  \emph{Computer Music Journal}, vol.~29, no.~2, pp. 61--76, 2005.

\bibitem{meredith2006ps13}
D.~Meredith, ``The ps13 pitch spelling algorithm,'' \emph{Journal of New Music
  Research}, vol.~35, no.~2, pp. 121--159, 2006.

\bibitem{longuet_higgins1971_bach}
H.~C. Longuet-Higgins and M.~Steedman, ``On interpreting {Bach},''
  \emph{Machine intelligence}, vol.~6, 1971.

\bibitem{wetherfield_2020_pitchspelling}
B.~Wetherfield, ``The minimum cut pitch spelling algorithm,'' in
  \emph{International Conference of Technologies for Music Notation and
  Representation (TENOR)}.\hskip 1em plus 0.5em minus 0.4em\relax Hamburg
  University for Music and Theater, 2020, pp. 149--157.

\bibitem{teodoru2007pitch}
G.~Teodoru and C.~Raphael, ``{Pitch Spelling with Conditionally Independent
  Voices.}'' in \emph{International Society for Music Information Retrieval
  Conference (ISMIR)}, 2007, pp. 201--206.

\bibitem{honingh2009lattice}
A.~K. Honingh, ``Compactness in the {Euler}-lattice: A parsimonious pitch
  spelling model,'' \emph{Musicae Scientiae}, vol.~13, no.~1, pp. 117--138,
  2009.

\bibitem{chew2002spiral}
E.~Chew, ``The spiral array: An algorithm for determining key boundaries,'' in
  \emph{International Conference on Music and Artificial Intelligence}.\hskip
  1em plus 0.5em minus 0.4em\relax Springer, 2002, pp. 18--31.

\bibitem{temperley2008key}
D.~Temperley and E.~W. Marvin, ``Pitch-class distribution and the
  identification of key,'' \emph{Music Perception}, vol.~25, no.~3, pp.
  193--212, 2008.

\bibitem{albrecht2013largecorpora}
J.~Albrecht and D.~Shanahan, ``The use of large corpora to train a new type of
  key-finding algorithm: An improved treatment of the minor mode,'' \emph{Music
  Perception: An Interdisciplinary Journal}, vol.~31, no.~1, pp. 59--67, 2013.

\bibitem{lopez2019key}
N.~N. L{\'o}pez, C.~Arthur, and I.~Fujinaga, ``{Key-finding based on a hidden
  Markov model and key profiles},'' in \emph{Digital Libraries for Musicology
  (DLfM)}, 2019, pp. 33--37.

\bibitem{foscarin2020asap}
F.~Foscarin, A.~Mcleod, P.~Rigaux, F.~Jacquemard, and M.~Sakai, ``{ASAP}: a
  dataset of aligned scores and performances for piano transcription,'' in
  \emph{International Society for Music Information Retrieval Conference
  (ISMIR)}, 2020.

\bibitem{SHERSTINSKY2020rnn}
A.~Sherstinsky, ``Fundamentals of recurrent neural network ({RNN}) and long
  short-term memory ({LSTM}) network,'' \emph{Physica D: Nonlinear Phenomena},
  vol. 404, pp. 132--306, 2020.

\bibitem{micchi2020not}
G.~Micchi, M.~Gotham, and M.~Giraud, ``Not all roads lead to {Rome}: Pitch
  representation and model architecture for automatic harmonic analysis,''
  \emph{Transactions of the International Society for Music Information
  Retrieval (TISMIR)}, vol.~3, no.~1, pp. 42--54, 2020.

\bibitem{oore2020time}
S.~Oore, I.~Simon, S.~Dieleman, D.~Eck, and K.~Simonyan, ``This time with
  feeling: Learning expressive musical performance,'' \emph{Neural Computing
  and Applications}, vol.~32, no.~4, pp. 955--967, 2020.

\bibitem{thickstun2018coupled}
J.~Thickstun, Z.~Harchaoui, D.~P. Foster, and S.~M. Kakade, ``Coupled recurrent
  models for polyphonic music composition,'' in \emph{International Society for
  Music Information Retrieval Conference (ISMIR)}, 2018.

\bibitem{midi_spec}
\BIBentryALTinterwordspacing
D.~Back, ``Standard {MIDI}-file format specifications,'' 1999, accessed May 5,
  2021. [Online]. Available:
  \url{http://www.music.mcgill.ca/~ich/classes/mumt306/StandardMIDIfileformat.html}
\BIBentrySTDinterwordspacing

\bibitem{gronlund2017fast}
A.~Gr{\o}nlund, K.~G. Larsen, A.~Mathiasen, J.~S. Nielsen, S.~Schneider, and
  M.~Song, ``Fast exact k-means, k-medians and {Bregman} divergence clustering
  in 1d,'' \emph{arXiv preprint arXiv:1701.07204}, 2017.

\bibitem{dowling1978scale}
W.~J. Dowling, ``Scale and contour: Two components of a theory of memory for
  melodies.'' \emph{Psychological review}, vol.~85, no.~4, p. 341, 1978.

\bibitem{chen2019harmony}
T.-P. Chen and L.~Su, ``Harmony transformer: Incorporating chord segmentation
  into harmony recognition,'' \emph{neural networks}, vol.~12, p.~15, 2019.

\bibitem{murphy2012machine}
K.~P. Murphy, \emph{Machine learning: a probabilistic perspective}.\hskip 1em
  plus 0.5em minus 0.4em\relax {MIT Press}, 2012.

\bibitem{collobert2008unified}
R.~Collobert and J.~Weston, ``A unified architecture for natural language
  processing: Deep neural networks with multitask learning,'' in
  \emph{International conference on Machine learning}, 2008, pp. 160--167.

\bibitem{deng2013new}
L.~Deng, G.~Hinton, and B.~Kingsbury, ``New types of deep neural network
  learning for speech recognition and related applications: An overview,'' in
  \emph{IEEE international conference on acoustics, speech and signal
  processing}, 2013, pp. 8599--8603.

\bibitem{kingma_adam:_2015}
D.~Kingma and J.~Ba, ``Adam: A method for stochastic optimization,'' in
  \emph{Proceedings of the International Conference on Learning Representations
  ({ICLR})}, 2015.

\bibitem{10.1162/neco.1997.9.8.1735}
S.~Hochreiter and J.~Schmidhuber, ``Long short-term memory,'' \emph{Neural
  Comput.}, vol.~9, no.~8, p. 1735–1780, Nov. 1997.

\bibitem{cho-etal-2014-learning}
K.~Cho, B.~van Merri{\"e}nboer, C.~Gulcehre, D.~Bahdanau, F.~Bougares,
  H.~Schwenk, and Y.~Bengio, ``Learning phrase representations using {RNN}
  encoder{--}decoder for statistical machine translation,'' in \emph{Conference
  on Empirical Methods in Natural Language Processing ({EMNLP})}, Doha, Qatar,
  Oct. 2014, pp. 1724--1734.

\bibitem{JMLR:v15:srivastava14a}
N.~Srivastava, G.~Hinton, A.~Krizhevsky, I.~Sutskever, and R.~Salakhutdinov,
  ``Dropout: A simple way to prevent neural networks from overfitting,''
  \emph{Journal of Machine Learning Research}, vol.~15, no.~56, pp. 1929--1958,
  2014.

\bibitem{zaheer2020bigbird}
M.~Zaheer, G.~Guruganesh, K.~A. Dubey, J.~Ainslie, C.~Alberti, S.~Ontanon,
  P.~Pham, A.~Ravula, Q.~Wang, L.~Yang \emph{et~al.}, ``Big bird: Transformers
  for longer sequences,'' \emph{Advances in Neural Information Processing
  Systems}, vol.~33, 2020.

\bibitem{xiong2021nystromformer}
Y.~Xiong, Z.~Zeng, R.~Chakraborty, M.~Tan, G.~Fung, Y.~Li, and V.~Singh,
  ``Nystr{\"o}mformer: A {Nystr{\"o}m}-based algorithm for approximating
  self-attention,'' 2021.

\end{thebibliography}

\end{document}